\documentclass{ws-mplb}

\begin{document}

\markboth{F. Sorrentino, G. Ferrari, N. Poli, R. Drullinger and G.
M. Tino}{Laser Cooling of Atomic Strontium}


%

\catchline{}{}{}{}{}

%


\title{LASER COOLING AND TRAPPING OF ATOMIC STRONTIUM FOR ULTRACOLD ATOMS
PHYSICS, HIGH-PRECISION SPECTROSCOPY AND QUANTUM SENSORS}

\author{\footnotesize F. Sorrentino, G. Ferrari, N. Poli, R. Drullinger\footnote{on leave from NIST, 325 Broadway, Boulder, Colorado 80305} and G. M. Tino}

\address{Dipartimento di Fisica and LENS - Universit\`a di Firenze, INFN,
INFM,\\ I-50019 Sesto Fiorentino (FI), Italy\\
guglielmo.tino@fi.infn.it}

\maketitle

\begin{history}

\received{(received date)}

\revised{(revised date)}



\end{history}

\begin{abstract}

This review describes the production of atomic strontium samples
at ultra-low temperature and at high phase-space density, and
their possible use for physical studies and applications. We
describe the process of loading a magneto-optical trap from an
atomic beam and preparing the sample for high precision
measurements. Particular emphasis is given to the applications of
ultracold Sr samples, spanning from optical frequency metrology to
force sensing at micrometer scale.
\end{abstract}

\section{Introduction}

\label{Introduction}

Recently, laser-cooled atomic strontium has been the subject of
active research in several fields spanning from all-optical
cooling towards quantum degeneracy for
bosonic,\cite{KatoriMOT1999,Ferrari2006} and
fermionic\cite{Mukaiyama2003} isotopes, cooling
physics,\cite{Xu2003,Loftus2004} continuous atom
laser,\cite{Katori2001} detection of ultra-narrow
transitions,\cite{Takamoto2003,Courtillotclock,Ferrari2003,Ido2005}
multiple scattering,\cite{Bidel2002} and collisional
theory\cite{Derevianko2003}.

Much of that interest relies on the features of the electronic
level structure of alkali-earth atoms, that make them ideal
systems for laser manipulation and for the realization of quantum
devices. Among the alkali-earth metals, strontium summarizes most
of the useful properties both for the preparation of ultracold
samples and for applications.

The ground-state Sr atom presents a strong and quasi-closed
optical transition well suited for efficient trapping in
magneto-optical traps (MOT's) from thermal
samples,\cite{Raab1987,Kurosu1990} and a multiplet of weak
intercombination transitions with large interest in laser cooling
and optical metrology. All of these transitions are easily
accessible with compact solid-state laser sources, as described in
\ref{ExperimentalSetup}. Second-stage cooling on the narrow
$^1$S$_0$-$^3$P$_1$ intercombination line was proven to be an
efficient method to reduce the sample temperature down to the
photon recoil limit.\cite{KatoriMOT1999}

The absence of a nuclear spin in all of the bosonic isotopes
greatly simplifies the electronic energy spectrum with respect to
the already simple structure of alkali atoms, and allows a direct
verification of theories in the field of light
scattering\cite{Chanelière2004} and laser
cooling.\cite{Loftus2004}  Moreover, the lack of nuclear spin
results in minimum sensitivity to stray magnetic fields: this has
important consequences on high-precision measurements, as
discussed in sections  \ref{Microsensor} and \ref{1S3PFreqMeasure}.

Interatomic collisions may represent an important source of
perturbations in atomic quantum devices. In this respect, the
$^{88}$Sr atom exhibits excellent features due to its remarkably
small collisional cross-section, resulting in the longest
coherence time for Bloch oscillations
observed so far.\cite{FerrariBloch}

As a consequence of their zero magnetic moment, ground-state
bosonic Sr isotopes cannot be magnetically trapped. However,
optical dipole trapping with far-off resonant laser fields
provides an effective method for trapping in conservative
potentials. By properly choosing the wavelength of the trapping
laser field, it is possible to design optical potentials that
shift equally levels belonging to the singlet and triplet
manifolds, hence avoiding perturbations to a given
intercombination line.\cite{KatoriFORT1999} This feature is
the basis of laser cooling in dipole traps and allows one to reach
a temperature close to the recoil limit with a phase-space density
close to quantum degeneracy both with bosonic and fermionic
isotopes.\cite{Ido2000,Mukaiyama2003,Ferrari2006}

Several groups are presently working on laser cooled strontium.
The Tokyo group first demonstrated the double-stage optical
cooling down to sub-$\mu$K temperatures;\cite{KatoriMOT1999} they
also proposed and realized a Sr ``optical lattice
clock'',\cite{Takamoto2005} and made pioneering studies on
electric microtraps for strontium.\cite{Katori2006}  The JILA
group has made interesting studies on cooling
physics,\cite{Xu2003,Loftus2004} and on metrological
applications.\cite{Ido2005} Our group is concerned both with
ultracold physics studies,\cite{Ferrari2006} and with applications
to optical metrology\cite{Ferrari2003} and quantum
devices.\cite{FerrariBloch} The BNM-SYRTE group is mainly
concerned with metrological
applications,\cite{Courtillotclock,Lemonde2005,Targat2006} the Houston
group with ultracold Sr physics,\cite{Nagel2005,Mickelson2005} while the
Nice group has studied multiple scattering of light from cold Sr
samples,\cite{Chanelière2004} and analyzed several details of the
Sr cooling mechanisms.\cite{Chanelière2005}

From the experimental point of view, some of the techniques and
solutions involved in atomic strontium cooling are specific to
this atom and to some extent uncommon in the field of atomic
physics and laser cooling. In this paper we give a detailed
description of how to prepare an ultra-cold strontium sample which
is well suited for high-precision spectroscopy, the study of
quantum degenerate gases and quantum sensors. The presentation has
the following structure:

\begin{itemize}
\item section \ref{Strontiumatom} summarizes the properties of the
strontium atom, \item section \ref{MOT} describes the process of
laser cooling of strontium starting from the slowing of a thermal
beam, to the cooling down to the photon recoil limit and trapping
in a conservative potential, \item section
\ref{CollisionalMeasure} deals with the study of ultracold
collisions on the ground-state bosonic Sr isotopes, \item in
section \ref{SrBEC} we show some recent advances towards the
realization of a Bose-Einstein condensate (BEC) of strontium,
\item in section \ref{Microsensor} we present the use of ultracold
Sr atoms as force sensors at micrometric scale, \item in section
\ref{1S3PFreqMeasure} we review the recent frequency measurements
on the Sr intercombination transitions with application to optical
metrology, \item in \ref{ExperimentalSetup} we give a detailed
description of the experimental setup employed in laser cooling of
strontium, namely the vacuum apparatus and the laser sources.
\end{itemize}

\section{Properties of the strontium atom}
\label{Strontiumatom}

The main properties of atomic strontium are common to almost all
alkaline-earth metals. At ambient temperature it appears as a
metal. Its vapor pressure is quite low, and it reaches 1 mTorr at
1000\,Kelvin. The Sr atom is rather reactive: it forms compounds
with oxygen, nitrogen, water and silicates, while it is inert
against sapphire. Thus in working with Sr vapors it is common to
employ sapphire windows.\cite{Neuman1995}

Strontium has four natural isotopes, whose properties are listed
in tab. 1. The bosonic (even) isotopes have zero nuclear spin,
thus they are perfect scalar particles in the $J=0$ states. This
has important consequences for applications to optical metrology
and quantum sensors (see sections \ref{Microsensor} and
\ref{1S3PFreqMeasure}).

\begin{table}[pt]
\tbl{Natural Sr isotopes (NIST data)} {\begin{tabular}{@{}cccc@{}}
\toprule Isotope & Relative atomic mass & Relative abundance &
Nuclear spin  \\ \colrule
$^{88}$Sr &   87.905 6143(24) & \hphantom{0}82.58(1)\% & 0 \\
$^{86}$Sr &  85.909 2624(24) & \hphantom{00}9.86(1)\% & 0 \\
$^{87}$Sr &  86.908 8793(24) & \hphantom{00}7.00(1)\% & 9/2 \\
$^{84}$Sr &  83.913 425(4) & \hphantom{000}0.56(1)\% & 0\\
\botrule
\end{tabular}}
\end{table}

\begin{figure}[t]
\centerline{\psfig{file=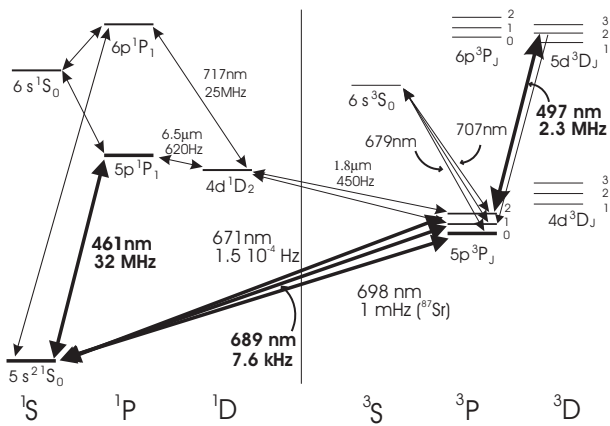,width=8cm}} \vspace*{8pt}
\caption{Electronic level structure of atomic strontium. The
transitions relevant for laser cooling and optical manipulation
are indicated as well as their linewidths.}\label{LevelScheme}
\end{figure}

Concerning the electronic level structure, due to the presence of
two electrons in the outer shell the atomic states can be grouped
into two separate classes: singlets and triplets. Since the
spin-orbit interaction breaks the spin symmetry, intercombination
transitions between singlet and triplet states are weakly
allowed.\cite{LibroFisicaAtomica} A simplified scheme of relevant
Sr levels and transitions is shown in figure \ref{LevelScheme}.

The strong $^1$S$_0$-$^1$P$_1$ transition at 461 nm has a natural
width of 32 MHz, and it is used for laser cooling and
trapping.\cite{Raab1987,Kurosu1990} Such transition is not
perfectly closed, due to a small leakage towards the 4d\,$^1$D$_2$
state (branching ratio $\sim 10^{-5}$). The direct
$^1$D$_2$-$^1$S$_0$ decay channel is forbidden in dipole
approximation, and atoms from the $^1$D$_2$ basically decay
towards the 5p\,$^3$P$_2$ (branching ratio 33\,\%) and
5p\,$^3$P$_1$ (branching ratio 67\,\%) states.

The line strength of the three 5s$^2$\,$^1$S-5s5p\,$^3$P
intercombination transitions range from the relatively high value
(7.6\,kHz) of the 689\,nm $^1$S$_0$-$^3$P$_1$ line to the
virtually zero value of the $^1$S$_0$-$^1$P$_0$ line for the
even isotopes in the absence of external fields. In $^{87}$Sr the
 0-0 line has a natural
linewidth of
$\sim 1$\,mHz due to hyperfine mixing.

\section{Laser cooling and trapping}
\label{MOT}

A magneto-optical trap operated on the
$^1$S$_0$-$^1$P$_1$ line requires the use of a blue laser source,
that can be realized through second-harmonic generation from a
semiconductor laser, as described in section \ref{BlueLaser}.

The final temperature in such a ``blue MOT'' is limited to few mK
by the linewidth of the $^1$S$_0$-$^1$P$_1$ transition: the ground
state of alkaline-earth even isotopes has no hyperfine or Zeeman structure, and this
prevents the application of sub-Doppler cooling techniques.
However, the presence of narrow intercombination transitions
allows efficient second-stage Doppler cooling down to sub-recoil
temperatures, as described in \ref{RedMOT}. In $^{87}$Sr, the
ground-state hyperfine structure offers the chance for sub-Doppler
cooling on the
$^1$S$_0$-$^1$P$_1$ transition, as demonstrated by the JILA
group.\cite{Xu2003} Alternatively, it is possible to apply sub-Doppler
cooling techniques to atoms trapped in the metastable $^3$P$_2$ level, as
already demonstrated on Ca.\cite{Grunert}

In this section we illustrate the Sr cooling and trapping in
detail, as performed in our laboratory. We start by preparing a mK
sample in the blue MOT from a Zeeman-slowed atomic beam, using optical pumping to recycle atoms
from the metastable $^3$P$_2$ level. Then we transfer the atoms to
a ``red MOT'' operated on the $^1$S$_0$-$^3$P$_1$ intercombination
line, where we cool them down to $\mu$K temperatures. Our
apparatus allows us to trap different Sr isotopes simultaneously,
as described in \ref{mixture}. The final step consists in loading
an optical dipole trap.

\subsection{Zeeman slowing and atomic beam collimation} \label{BeamCollimation}

The sequence for cooling and trapping on the 461\,nm resonance
line follows standard techniques, as already reported by other
groups.\cite{KatoriMOT1999,Xu2003,Kurosu1990,Oates1999} We slow
the thermal atomic beam to few tens of m/s in a 30-cm long
Zeeman-slower\cite{Prodan1982} based on a two-stages tapered coil
with a zero crossing magnetic field, and a counter-propagating
laser beam frequency shifted by 480\,MHz to the red of the
$^1$S$_0$-$^1$P$_1$ transition. The beam, with typical power of
40\,mW, has a 7\,mm $1/e^2$ radius at the MOT and it is focused on
the capillaries (see appendix \ref{VacuumSystem}). The distance
between the capillaries and the MOT region is 100\,cm.

The atomic beam brightness can be increased with a 2-D transverse
cooling stage before the Zeeman slower.\cite{Shimizu,Rasel1999}
This improves the MOT loading both by increasing the atomic flux
coupled into the differential pumping tube (see
\ref{VacuumSystem}), and by effectively reducing the final
diameter of the atomic beam after the Zeeman slower.

We implement a 2D optical molasses with two 461\,nm beams sent
perpendicularly to the atomic beam, in multipass geometry on two
orthogonal planes. The beams are red detuned by 20\,MHz with
respect to the $^1$S$_0$-$^1$P$_1$ resonance. At given optical power, the
multi-pass geometry improves the transverse cooling efficiency  by
increasing the length of the interaction region. In the present setup we
use beams with
$1/e^2$ diameter of 2.5\,mm$^2$ bouncing about 12 times to cover an
interaction length of 4\,cm. The interaction starts 15\,cm after the
capillaries (see
\ref{VacuumSystem}). In that region the beam diameter has already
reached about 11\,mm.

Without beam collimation we can couple into the differential
pumping tube (placed 24\,cm far from collimation region) about
10\,\% of the total atomic flux. With the 2D molasses we increase
the coupling by a factor of 4 in optimal conditions (laser detuning
$\sim 20$\,MHz, optical power $\sim 20$\,mW). This simple scheme can also
be used to deflect the atomic beam.

\subsection{Blue MOT} \label{BlueMOT}

The slowed atoms are then cooled and trapped in a MOT operated
 on the $^1$S$_0$-$^1$P$_1$ transition 40\,MHz to the red
with respect to resonance. Three retro-reflected beams with a
5\,mm $1/e^2$ radius cross in the MOT region with almost mutually
orthogonal directions. The vertical beam is collinear with the
magnetic quadrupole axis of an anti-Helmholz pair of coils. The
field gradient at the quadrupole center is 50\,Gauss/cm. Taking
into account the coupling efficiency of the blue laser source into
the fiber (70\,\%) and the efficiency of the AOMs ($75 \div
80$\,\%), the remaining 120\,mW are split into the three channels:
typical power values are respectively 60\,mW for MOT beams, 40\,mW
for Zeeman slower and 20\,mW for transverse cooling beams. The
total 461\,nm light incident on the MOT region amounts to $\sim
90$\,mW/cm$^2$.

\begin{figure}[t]
\centerline{\psfig{file=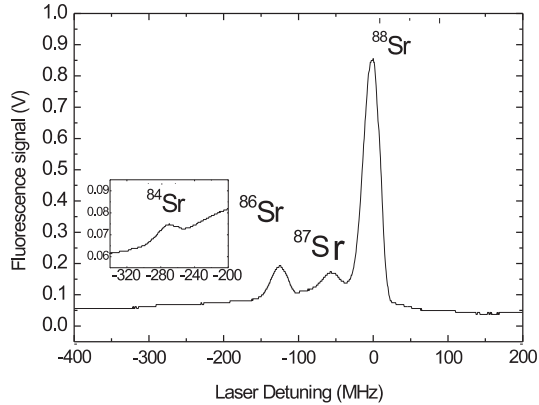,width=7cm}} 
\caption{Fluorescence of blue MOT as a function of laser detuning.
Resonances corresponding to the different isotopes are visible.
}\label{IsotopeFluorescence}
\end{figure}

We have a rough estimate of the number of trapped atoms by
collecting the 461\,nm fluorescence on a calibrated photodiode. By
changing the blue laser detuning we are able to separately trap
the four natural Sr isotopes, as shown in figure
\ref{IsotopeFluorescence}. For a more accurate measure of the atom
number we perform absorption imaging on a CCD with a resonant
461\,nm horizontally propagating probe beam. We measure the sample
temperature with standard time-of-flight technique.

The lowest temperatures measured in the blue MOT are typically
higher than the Doppler limit by at least a factor two. The Nice
group has shown that the extra-heating mechanisms causing such
discrepancy can be explained in terms of intensity fluctuations in the MOT
laser beams.\cite{Chanelière2005}

\subsection{Optical repumping from metastable states} \label{OpticalPumping}

The $^1$S$_0$-$^1$P$_1$ transition used in the first stages of
cooling and trapping is not perfectly closed due to the decay
channel of the 5s5p\,$^1$P$_1$ towards the 5s4d\,$^1$D$_2$ state,
that has a branching ratio of $2 \times 10^{-5}$. Atoms in the
latter state may decay back to the ground state trough the
5s5p\,$^3$P$_1$ within less than 1 \,ms, or may decay to the
metastable 5s5p\,$^3$P$_2$ and be lost (see figure
\ref{LevelScheme}). Under typical MOT conditions this process
limits the MOT lifetime to few tens of milliseconds.

\begin{figure}[t]
\centerline{\psfig{file=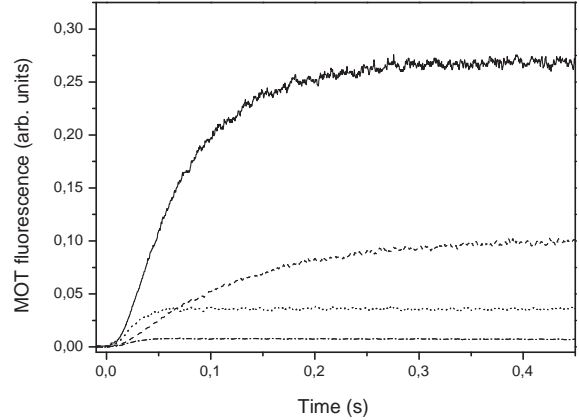,width=9cm}} 
\caption{Fluorescence of the charging MOT. Solid line: with
transverse cooling and repumper. Dash: with repumper. Dot: with
transverse cooling. Dash-dot: without repumper or transverse
cooling.}\label{BlueMOTFluorescence}
\end{figure}

Some groups already circumvented this limitation by optical
pumping atoms from the metastable 5s5p\,$^3$P$_2$ to the ground
state via the 5s6s\,$^3$S$_1$ state with 707\,nm
light.\cite{Vogel1999} Since the 5s6s\,$^3$S$_1$ state is also
coupled to the metastable 5s5p\,$^3$P$_0$ an additional laser at
679 nm is necessary in this scheme. To reduce the number of
repumping lasers we use a different approach which involves the
5s5d\,$^3$D$_2$ state and requires one single laser at 497 nm.
Concerning the optical pumping, the $^3$D$_2$ state is coupled
basically to the $^3$P$_2$ and $^3$P$_1$ states, then efficient
pumping is insured within few absorption cycles. To this purpose
during the loading we illuminate the blue MOT with light at
497\,nm, produced by the source described in section
\ref{BlueLaser}, which is kept on resonance with the repumping
transition. Figure \ref{BlueMOTFluorescence} shows the effect of
the repumping field on the $^{88}$Sr MOT loading in different
conditions. Without the repumper the MOT loading time ($1/e$) is
10.2 ms regardless to the presence of the atomic beam transverse
cooling, and correspondingly to the flux of atoms captured. In
presence of the repumper we observe a 0.24 s lifetime when few
atoms are present in the MOT, and a decreasing lifetime down to
0.11 s at full MOT charging. This reduction in lifetime is
explained in terms of light assisted collision.\cite{Vogel1999}

Due to the difference in the natural isotope abundance (see
section \ref{Strontiumatom}), the loading flux into the MOT varies
correspondingly. In typical conditions, i.e. when operating with
the transverse cooling and optical repumping from the metastable
state, and with the Sr oven kept at $\sim 700$\,K, the
steady-state blue MOT population amounts to $\sim 10^8$ atoms for
$^{88}$Sr and $\sim 10^7$ atoms for $^{86}$Sr.

\subsection{Red MOT} \label{RedMOT}

The Tokyo group first realized a strontium MOT operated on the
$^1$S$_0$-$^3$P$_1$ line.\cite{KatoriMOT1999} Such system has been
exhaustively studied by the JILA group, both theoretically and
experimentally.\cite{Loftus2004} The dynamics of laser cooling on
narrow transitions presents several interesting features. Unlike
in the case of ordinary magneto-optical traps, for the
$^1$S$_0$-$^3$P$_1$ transition in alkaline-earth atoms the natural
linewidth $\Gamma$ is of the order of the recoil frequency shift
$\Gamma_R$. In such conditions both mechanical and thermodynamical
MOT properties cannot be explained by the ordinary semiclassical
Doppler theory of laser cooling. In particular, the role of
gravity becomes non negligible, and the atomic temperature can be
lower than the recoil limit $T_R = \frac{\hbar \Gamma_R}{k_B}$.

\begin{figure}[t]
\centerline{\psfig{file=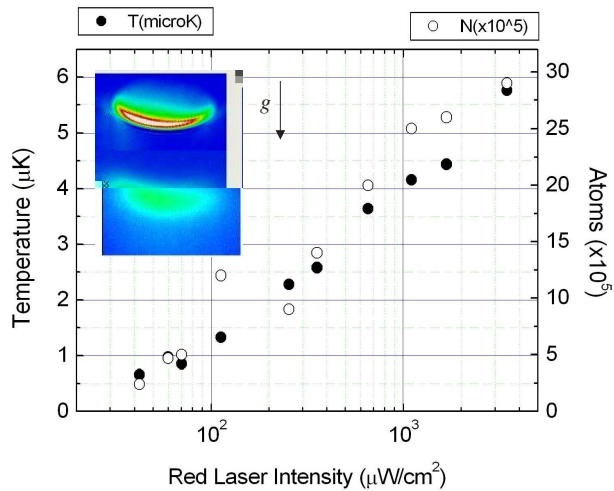,width=9cm}} 
\caption{Measured atomic temperature and population of the red MOT
as a function of the laser intensity with our apparatus. The inset
shows in-situ and free-fall images of the trapped
atoms.}\label{RedMOTtemperature}
\end{figure}

When the laser detuning $\delta$ is negative and larger than the
power-broadened linewidth $\Gamma\sqrt{1+\frac{I}{I_{sat}}}$, the
atoms interact with the MOT laser beams only on a thin ellipsoidal
shell, corresponding to the surface where the laser frequency
offset compensates for the Zeeman shift. The maximum radiation
force is only one order of magnitude larger than gravity. As a
result, the atoms sag on the bottom of the ellipsoid, as
shown in the inset of figure \ref{RedMOTtemperature}. In such
conditions the atomic temperature is mainly determined by the
interaction with the upward propagating vertical MOT beam. With
typical laser intensities of $I \simeq 10^1 \div 10^4 I_{sat}$,
measured temperatures are in good agreement with a semiclassical
model, and can be expressed as

\begin{equation}
T=\frac{\hbar \Gamma \sqrt{1+\frac{I}{I_{sat}}}}{2 k_B},
\end{equation}
which is lower than the Doppler temperature at the same detuning.
At low laser intensity ($I \approx I_{sat}$), the cooling
mechanism becomes fully quantum-mechanical, and the minimal
attainable temperature is $T_R \over 2$.\cite{Loftus2004} Figure
\ref{RedMOTtemperature} shows the dependence of the MOT
temperature on the 689\,nm laser intensity, as measured in our
laboratory.

The JILA group also studied the case of positive laser detuning,
revealing the appearance of novel striking phenomena such as
momentum-space crystals. They showed that driving the atomic
system with 689\,nm light blue detuned from the
$^1$S$_0$-$^3$P$_1$ resonance may result in a periodic pattern in
the atomic velocity distribution.\cite{Loftus2004}

Second-stage cooling of the odd Sr isotope is more complex,
due to the hyperfine structure of both ground and excited states.
The Tokyo group has shown that using two lasers at 689\,nm
it is possible to cool and trap $^{87}$Sr atoms at phase-space
densities close to the Fermi degeneracy.\cite{KatoriFermion}

We here illustrate the red MOT for bosonic Sr isotopes as realized
in our laboratory. The 689\,nm light for the MOT beam (14\,mW) is
provided by a slave laser injection-locked with light coming from
the stable master laser described in \ref{RedLaser}. The beam spot
is then enlarged to the same radius as the 461\,nm MOT beams and
overlapped to the blue beams with a dichroic mirror. After that
mirror, the two wavelengths share the same broad-band optics:
polarizing beam splitters, mirrors and waveplates.

The linewidth of the intercombination transition is not sufficient
to cover the Doppler broadening corresponding to the velocity
distribution of the sample trapped in the blue MOT. An efficient
solution consists in broadening the spectrum of the 689\,nm laser
field.\cite{KatoriMOT1999} On the contrary, groups working on
other alkaline-earth atoms (i.e. Ca and Mg) employ a resonant
coupling to some higher level to quench the $^3$P$_1$
lifetime,\cite{Binnewies2001} thus increasing the effective
strength of the intercombination transition, as the maximum
radiation force would be otherwise lower than gravity.

We add a frequency modulation at 50\,kHz in the first 200\,ms. The
total frequency span is 2\,MHz, corresponding to 40 sidebands,
with an intensity of 120\,$\mu$W/cm$^2$ per sideband (the
saturation intensity is $I_{sat} = 3$\,$\mu$W/cm$^2$). The central
frequency is red detuned by 1 MHz with respect to resonance. At
the end of the FM recapture, we normally obtain a cloud at a
temperature of about 20\,$\mu$K. After that we remove the
frequency modulation, set a fixed red detuning from
$^1$S$_0$-$^3$P$_1$ transition and reduce the beam intensity in
the last 60\,ms of cooling. Working at 350\,kHz below resonance
and reducing the total light intensity on the MOT to
70\,$\mu$W/cm$^2$, we can then transfer about 10\,\% of the atoms
from the blue MOT to the red MOT at temperatures below 1\,$\mu$K.

\subsection{Cooling and trapping isotopic mixtures} \label{mixture}

Among the experiments on ultracold atoms, much work is being
concentrated on the study of mixtures of different atomic
species\cite{Santos1995,Modugno2001,Murdrich2002} or different
isotopes of the same
species.\cite{Mewesl999,Loftus2001,Honda2002,Stas2004} Mixtures
offer a way to exploit collisional physics not applicable in
single species samples.\cite{Schreck2001} They also offer
additional degrees of freedom, such as sympathetic cooling, in
order to achieve the degenerate quantum regime with atoms for
which evaporative cooling is not efficient.\cite{Khaykovich2001}

\begin{figure}[t]
\centerline{\psfig{file=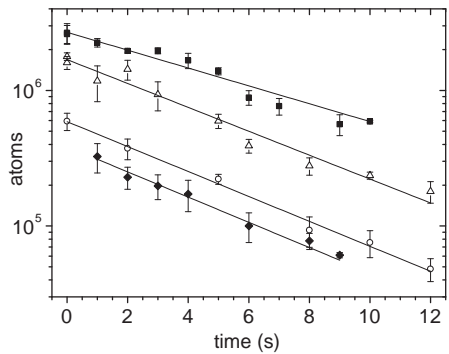,width=7cm}} 
\caption{Decay of the $^3$P$_2$ state when trapped in a 56\,G/cm magnetic quadrupole. Circles
($^{86}$Sr) and squares $^{88}$Sr refer  to data taken for the individual isotopes with the blue
MOT switched off; diamonds ($^{86}$Sr) and triangles ($^{88}$Sr) with the blue MOT operating on
the undetected isotope. The measurements are taken after red MOT recapture. Reprinted figure with
permission from N. Poli et al., Phys. Rev. A 71, 061403(R) (2005). Copyright (2005) by the
American Physical Society.}\label{BlueIsotopeLifetime}
\end{figure}

For simultaneous trapping of different isotopes previous
experiments employed laser sources delivering the necessary
frequency components for each isotope
involved.\cite{Suptitz1994,Mewesl999} In the case of the strontium
blue MOT, this approach may be difficult to apply because of the
complexity of the laser sources and the limited laser power. An
alterative solution is presented by the use of the magnetically
trapped $^3$P$_2$  state as a dark atom
reservoir.\cite{Stuhler2001,Nagel2003,Poli2005} During the blue
MOT phase without the repumper, the small loss channel of the
excited $^1$P$_1$ state towards the metastable $^3$P$_2$ state
provides a continuous loading of atoms into the potential given by
the MOT's magnetic quadrupole. Figure \ref{BlueIsotopeLifetime}
reports our lifetime measurements on the magnetically trapped
metastable isotopes. By using the same blue laser source, one can
sequentially load different isotopes into the magnetic potential
by just stepping the laser frequency to the different resonances.

We typically start by collecting one isotope (say $^{86}$Sr) for a
few seconds, then we tune the blue laser on resonance to the other
isotope (say $^{88}$Sr). Once the isotopic mixture is prepared in
the $^3$P$_2$ state, the blue light is switched off, and the FM
red MOT is switched on as well as the repumping beam. The isotopic
shift on the repumping transition is smaller than the resonance
width of the $^3$P$_2$-$^3$D$_2$ transition observed on the blue
MOT fluorescence, which results in efficient, simultaneous optical
pumping of $^{88}$Sr and $^{86}$Sr on a time scale short with
respect to the red MOT capture time. The loading of a single
isotope into the magnetic potential was already described by Nagel
et al.,\cite{Nagel2003}  and we did not observe significant
differences in the behavior when loading two isotopes. Figure
\ref{BlueIsotopeLifetime} shows the measurement of the lifetime
for each isotope, both when individually trapped and in the
presence of the blue MOT working on the other isotope. The
measured lifetime values are all of the order of 5\,s, close to
the background pressure limited lifetime of 7\,s.

The laser source for the operation of the two-isotope red MOT is
composed of two slave lasers injected from the same frequency
stabilized master with a frequency offset corresponding to the
isotopic shift of 163 817.3\,kHz.\cite{Ferrari2003} Subsequently,
the frequency and intensity of the two beams are controlled by
double pass AOMs driven by the same RF, the beams are superimposed
on a polarizing beam splitter, and then they are overlapped to the
blue MOT beams as described previously. Comparing the two-isotope
red MOT with the single isotope one, with the same atom numbers we
do not observe any effect in the transfer efficiency and final
temperature due to the presence of the second isotope. In this
way, we obtain samples with up to 10$^7$ (10$^6$) atoms of
$^{88}$Sr ($^{86}$Sr) at a temperature 2\,$\mu$K (1\,$\mu$K). We
attribute the difference in the loading to the natural abundances
and to minor differences in the red MOT parameters. By varying the
order of loading and the loading times of the two isotopes we can
vary almost arbitrarily the final ratio of populations.

The atom number and temperature are measured independently on the
two isotopes by absorption imaging with the resonant 461\,nm probe
beam, and the contribution of the non-resonant isotope is taken
into account.

\subsection{Optical dipole trap} \label{FORT}

In magneto-optical traps the atomic temperature and lifetime are
fundamentally limited by resonant photon scattering and
light-assisted collisions. In all experiments requiring long
storage times or ultra-low temperatures, it is convenient to
transfer the atoms into a conservative trap. The ground-state even
isotopes of alkaline-earth atoms cannot be magnetically trapped,
due to the absence of Zeeman structure. Though the Tokyo group has
recently demostrated a clever scheme for Sr trapping with AC elecric
fields,\cite{Katori2006}  in most cases the best choice consists in
optical dipole traps. Moreover on two-electrons atoms it is possible to
apply light-shift cancellation techniques to the intercombination
transitions,\cite{KatoriFORT1999} opening the way to optical spectroscopy
with ultimate accuracy. However, the optical dipole trap is widely
employed with magnetic atoms as well,\cite{opticaltrapping} since it
generally produces a stronger confinement than magnetic traps, up to the
Lamb-Dicke regime in optical lattices,\cite{Dicke} and permits trapping
in all the magnetic sub-levels.

In an optical dipole trap the confining force originates from the
energy level gradient produced by the intensity-dependent light
shift. The energy level shift $\Delta E$ of an atom in an optical
field is proportional to the light intensity $I$ and, in the rotating-wave approximation, inversely
proportional to the frequency detuning $\delta$ from resonance:

\begin{equation}
\Delta E = \frac{\hbar \Gamma^2 I}{8 \delta I_{sat}},
\end{equation}
where $\Gamma$ and $I_{sat}$ are the linewidth and the saturation
intensity of the resonance transition, respectively, while the
photon scattering rate $R_S$ scales as the light intensity and the
inverse square of the laser detuning:

\begin{equation}
R_S = \frac{\Gamma^3 I}{4 \delta^2 I_{sat}}.
\end{equation}
Thus, at a given trap potential depth, it is possible to reduce
the heating due to photon scattering by increasing $\delta$ and
$I$ proportionally. In far-off resonant optical dipole traps
(FORT) the effect of photon scattering is negligible for most
practical purposes.

The trapping radiation couples singlet and
triplet states independently. This results in a different
dependence on the trapping wavelength for the light shift of the
$^1$S$_0$ and $^3$P$_1$ levels, so that the differential shift
vanishes at a ``magic wavelength''. This enables optical cooling
into the dipole trap, since the detuning of the cooling radiation
is not position-dependent.

For our optical dipole trap we employ the apparatus described in appendix
\ref{IRlaser}. In most cases we use the horizontally propagating
beam alone. This single-beam FORT has a trap depth of $90\,\mu$K.
The computed radial and axial trap frequencies are $\omega_r =
2\pi \times 2$ kHz and $\omega_a = 2\pi \times 26$ Hz
respectively. The vertical beam produces a maximum potential depth
of about $12\,\mu$K. We find good agreement between computed and
measured trap frequencies. The resonant 461\,nm probe beam for
time-of-flight absorption imaging propagates horizontally at an
angle of about $30^{\circ}$ with the horizontal FORT beam.

\begin{figure}[t]
\centerline{\psfig{file=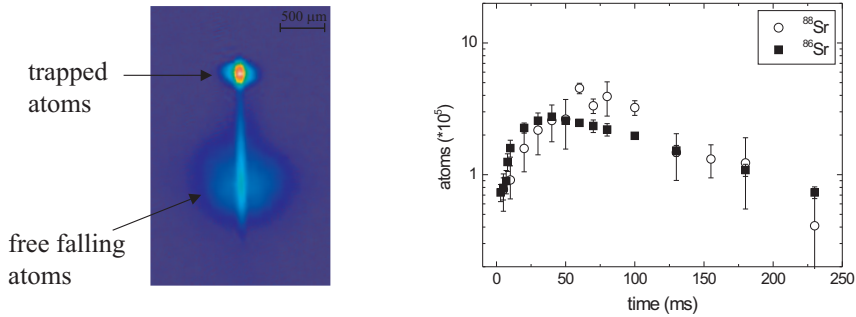,width=12cm}} 
\caption{Crossed-beams FORT loading. The inset shows an in-situ
image of trapped atoms. The graph shows the measured FORT
population as a function of the time overlap between FORT and red
MOT. Reprinted figure with permission from N. Poli et al., Phys.
Rev. A 71, 061403(R) (2005). Copyright (2005) by the American
Physical Society.}\label{DipoleTrap}
\end{figure}

The transfer efficiency from the red MOT to the FORT results from
a balance between loading flux and density dependent losses due to
light assisted collisions.\cite{KatoriFORT1999} Figure
\ref{DipoleTrap} shows typical loading curves, together with an
in-situ image of the trapped atoms.

The AC Stark shift of the intercombination line depends on the
direction of the FORT field with respect to the bias magnetic field
$\vec B$. We use the MOT quadrupole field to resolve the Zeeman structure
of the $^1$S$_0$-$^3$P$_1$ line, and we keep the polarization of the FORT
beam linear and orthogonal to
$\vec B$. In such conditions the resulting light shift is not critical
for laser cooling in the FORT. The wavelength used for the dipole trap is
only 7\,nm away from the ``magic wavelength'' for the intercombination
$^1$S$_0$-$^3$P$_1$ transition.\cite{Ido2003} When the polarization of
the dipole trapping light is orthogonal to the magnetic field the Stark
shift for the $^1$S$_0$-$^3$P$_1$ transition is lower than 10\,kHz.
Thus, it is possible to cool the atoms while loading the
dipole trap.\cite{Ido2000}

\section{Collisional studies on ground-state even Sr isotopes}
\label{CollisionalMeasure}

The study of atomic collisions at low temperature has undergone a
rapid development in recent years following the advent of a number
of cooling techniques for diluite atomic gases. The measurement of
collisional parameters gives insight into an interesting and
promising physics, allowing tests of theoretical models for
interatomic potentials and molecular
wavefunctions.\cite{Vogel1999,Fedichev1996} Moreover, a precise
knowledge of collisional properties is essential for experiments
aiming to explore regimes of quantum degeneracy in atomic
gases.\cite{Delannoy2001,Yamashita2003}

Here we describe a systematic analysis on the ground-state collisional
properties of the two most abundant bosonic Sr isotopes ($^{88}$Sr
and $^{86}$Sr) performed with our apparatus. More specifically, we
evaluated the elastic cross-sections $\sigma_{i-j}$ ($i,j =86,88$)
for both intra and inter-species collisions, and the three-body
recombination coefficients $K_{i}$ ($i =86,88$) in the FORT. The
elastic cross-sections were deduced by driving the system out of
thermal equilibrium and measuring the thermalization rate together
with the sample density. For the inelastic collisions, we measured
the density dependence of the trap loss rate. The key point of our
experimental procedure is a precise knowledge of the atomic
density. This in turn requires a proper control of the trap
frequencies. We assumed the equilibrium phase-space distribution
in an ideal harmonic trap to infer the atomic density from the
measured number of atoms $N$ and temperature $T$. Our assumption
is supported by the fact that the ratio $\eta=\frac{U}{k_B T}$ of
trap depth and sample temperature was larger than 5 in all of our
measurements. The equilibrium peak atomic density in a harmonic trap is

\begin{equation}
n_0 = N {\bar \nu}^3 \left({2 \pi m \over k_{B}T}\right)^{3 \over
2}
\end{equation}
where $\bar \nu$ is the average trap frequency, $m$ is the atomic
mass and $k_B$ is the Boltzmann constant. Our results show
significant differences in the collisional properties of the two
isotopes. Both the elastic cross-section and the three-body
collision coefficient were found to be several orders of magnitude
larger in $^{86}$Sr than in $^{88}$Sr, and the inter-species
cross-section $\sigma_{86-88}$ is much larger than the
intra-species cross-section $\sigma_{88-88}$.

We adopted the standard technique of observing the
cross-thermalization between orthogonal degrees of
freedom.\cite{Monroe1993,Arndt1997,Hopkins2000,Schmidt2003} The
single-beam FORT geometry is not well suited for this experiment,
as the temperature in the horizontal modes cannot be univocally
determined with our apparatus, and because with $^{86}$Sr in our
experimental conditions the system turned out to be in
hydrodinamic regime along the axial
direction.\cite{Odelin1999,Gensemer2001} Given these constraints,
we chose the crossed-beams geometry for the thermalization
measurements.

In order to drive the system out of thermal equilibrium we
performed a selective optical cooling along the vertical direction
with resonant 689 nm light. For this purpose we took advantage of
the fact that the FORT is located at the center of the red MOT,
where atoms are not resonant with the horizontal MOT
beams.\cite{Loftus2004} We chose the optical intensity and cooling
time in such way to create a detectable thermal anisotropy without
introducing dramatic losses.

\begin{figure}[t]
\centerline{\psfig{file=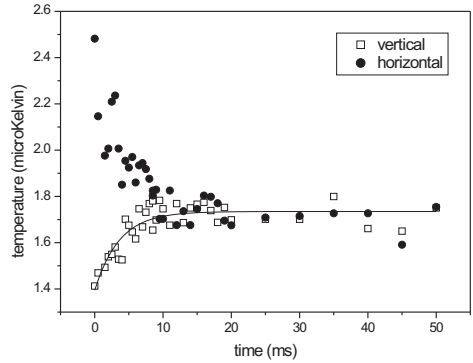,width=7cm}} 
\caption{Measurement of the thermalization rate for $^{86}$Sr. The
graph shows the temporal evolution of the horizontal and vertical
temperatures after cooling on the vertical direction. The solid
curve is an exponential fit to the vertical temperature
data.}\label{86Thermalization}
\end{figure}

We measured the intra-species thermalization rate $\tau_{th}$ by
separately loading one isotope (either  $^{88}$Sr or $^{86}$Sr) and
observing the temporal evolution towards thermal equilibrium (see
figure \ref{86Thermalization}). In both cases the total trap loss
rate was much lower than the thermalization rate, so the atom
number was constant within shot-to-shot fluctuations.

We deduced the elastic cross-section from the measured
thermalization rate, by assuming a value of 2.7 for the average
number of collisions to reach thermalization, that is,
$\tau_{th}=2.7
\tau_{coll}$.\cite{Monroe1993,Arndt1997,Kavoulakis2000} The
collision rate is given by

\begin{equation}
{1\over\tau_{coll}} = \bar n \sigma \bar v
\end{equation}
where $\bar n$ is the average atom density, $\bar v$ is the
average relative velocity of colliding atoms, and $\sigma$ is the
cross-section. We computed $\bar n$ and $\bar v$ from the measured
values for atom number $N$, average sample temperature $T$ and
average trap frequency $\bar \nu$. With our assumptions the
average density is given by $\bar n = {n_0 / 2 \sqrt 2}$, while
the average relative velocity is given by $\bar v = 4 \sqrt {k_B
T/\pi m}$.

We repeated our measurements for different values of the atom
density, in order to check whether the observed
cross-thermalization was due to elastic collisions or ergodic
mixing between different degrees of freedom. The resulting values
are $\sigma_{88-88} = 3(1) \times 10^{-13}$ cm$^2$  and
$\sigma_{86-86} = 1.3(0.5) \times 10^{-10}$ cm$^2$. The
uncertainty is mainly due to shot-to-shot fluctuations in the FORT
population, that reflect in both density and temperature
instabilities. Such results might be compared with the
zero-temperature cross sections deduced from scattering length
values through the relation $\sigma=8\pi\lambda^2$, where
$\lambda$ is the $s$-wave scattering length. Scattering length
values were obtained from photoassociation spectra by the Tokyo
group for $^{88}$Sr,\cite{yasuda2004} and by the Houston group for
both isotopes.\cite{Mickelson2005} As concerning $\sigma_{88-88}$,
our value is consistent with the Tokyo work, while the Houston
group predicts an even smaller cross-section. On the contrary,
there is a fair agreement between our $\sigma_{86-86}$ value and
the Houston work. All of these results are summarized in tab. 2

\begin{table}[pt]
\tbl{Intra-species elastic cross-section for $^{88}$Sr and
$^{86}$Sr, in cm$^2$} {\begin{tabular}{@{}cccc@{}} \toprule
Reference &
$\sigma_{88-88}$  & $\sigma_{86-86}$  & Method  \\
\colrule
Tokyo group\cite{yasuda2004} &  \, $3(1) \times 10^{-13}$ &  & $\lambda$ from PA spectra \\
Houston group\cite{Mickelson2005} &  $<1.2\times 10^{-13}$ & $2.6 \times 10^{-10} < \sigma < 3.7 \times 10^{-9}$ & $\lambda$ from PA spectra \\
Our group\cite{Ferrari2006} & \, $3(1) \times 10^{-13}$ & $1.3(0.5) \times 10^{-10}$ & cross-thermalization \\
\botrule
\end{tabular}}
\end{table}

\begin{figure}[t]
\centerline{\psfig{file=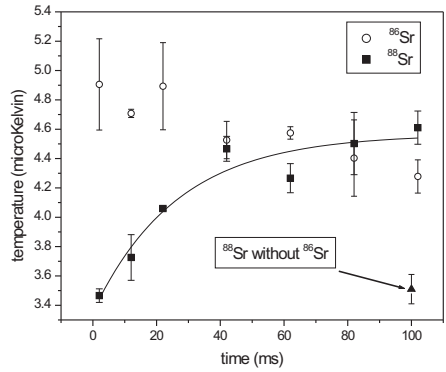,width=7cm}} 
\caption{Measurement of the interspecies thermalization rate. The
graph shows the temperature evolution of $^{88}$Sr and $^{86}$Sr
after selective cooling on $^{88}$Sr. The solid curve is an
exponential fit to the $^{88}$Sr data. Trap populations are $10^5$
atoms for $^{88}$Sr and $4.5 \times 10^4$ atoms for
$^{86}$Sr.}\label{MixThermalization}
\end{figure}

We adopted a similar approach for the measurement of the
inter-species cross section.\cite{Goldwin,K-Rb,Mosk,Delannoy} The
single-beam FORT was loaded with a mixture of $^{88}$Sr ($10^5$
atoms) and $^{86}$Sr ($4 \times 10^4$ atoms). We then applied a
selective optical cooling stage (duration 5\,ms) to $^{88}$Sr and
observed the temperature evolution of the two samples (see figure
\ref{MixThermalization}). The $^{88}$Sr is heated by $^{86}$Sr to
the equilibrium temperature in few tens of ms. For comparison, in
absence of $^{86}$Sr the temperature of $^{88}$Sr grows by less
than 5\,\% in 100\,ms. The resulting inter-species cross-section
was found as $\sigma_{88-86} = 4(1) \times 10^{-12}$ cm$^2$. This
value is significantly larger than the intra-species cross-section
$\sigma_{88-88}$, suggesting the way to a novel and efficient
sympathetic cooling mechanism.\cite{Ferrari2006}

We studied the inelastic collisions by loading a single isotope in
the single-beam FORT  (either  $^{88}$Sr or $^{86}$Sr) and looking
at the evolution of the number and temperature of trapped atoms in
the FORT after the MOT operation was finished. We found no
evidence for non-exponential decay with $^{88}$Sr. The measured
lifetime of 7\,s is consistent with the residual background gas
pressure of $10^{-8}$\,torr. With the initial atom density at trap
center of $3 \times 10^{13}$ cm$^{-3}$ this gives an upper limit
of $10^{-27}$ cm$^6$s$^{-1}$ for the $K_{88}$ coefficient.

\begin{figure}[t]
\centerline{\psfig{file=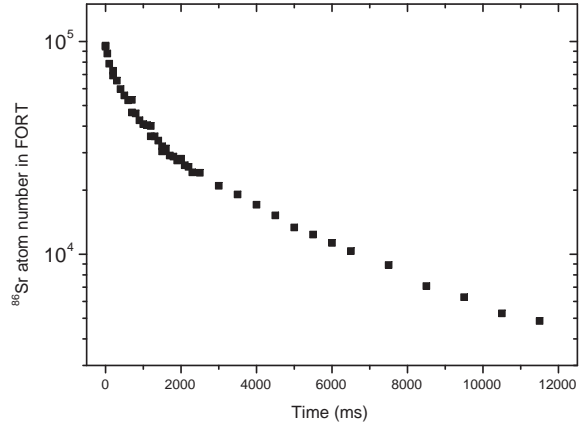,width=9cm}} 
\caption{Decay of trap population when loading $^{86}$Sr atoms
alone in the single-beam FORT. The sample temperature is
12\,$\mu$K, trap depth is 90\,$\mu$K.}\label{Recombination86}
\end{figure}

When loading the single-beam FORT with $^{86}$Sr only, we observed
a clear non-exponential decay (figure \ref{Recombination86}). We
deduced the three-body recombination coefficient from the density
dependence of the loss rate.\cite{Burt} Integrating the loss rate
equation

\begin{equation}
\dot N = -\Gamma_b N - K_{86}{\int {n^3 (\vec r, t) d^3 \vec r}}
\end{equation}
where $\Gamma_b$ is the linear loss rate for background gas
collisions, with our assumptions we can write

\begin{equation}
\ln {N \over N_0} = -\Gamma_b t - {K_{86} (2\pi)^3 m^3 \bar \nu ^6
\over  3^{3\over 2}}{\int {{N^2 (t) \over [k_B T(t)]^3} dt}}
\end{equation}
This formula is valid as long as additional losses due to
evaporation can be neglected. In order to limit such effect, we
select the coldest atoms by means of forced evaporation before the
measurement.

We deduced the linear loss rate from the wing of figure
\ref{Recombination86}. This value is consistent with the loss rate
measured with $^{88}$Sr. Then we performed a linear fit of $\ln {N
\over N_0} + \Gamma_b t$ versus $\int {{N^2 (t) \over [k_B
T(t)]^3} dt}$ to derive the recombination constant. We repeated
the measurement several times to average out density fluctuations
reflecting in large uncertainty on $K_{86}$. The final result was
$K_{86} = 1.0(0.5) \times 10^{-24}$ cm$^6$s$^{-1}$.

\section{Towards a BEC of strontium} \label{SrBEC}

Laser cooling is a very effective technique to reach phase-space
densities within few orders of magnitude from quantum degeneracy.
The limits in cooling at high density are set by the optical depth
of the sample, i.e. reabsorption of scattered light, and
light-assisted atom-atom collisions. Forced evaporative cooling
represents the common way to circumvent these
limits.\cite{Hess1986} However, this procedure is not effective
with all atoms. In particular, among the atoms cooled with optical
methods, none of the alkali-earth atoms reached quantum degeneracy
so far, except ytterbium which has an alkali-earth-like electronic
structure.\cite{Takasu2003} A phase-space density of $\simeq
10^{-1}$ was reported for Sr but BEC could not be
reached.\cite{Ido2000}

On this respect, the results of the collisional measurements
reported in \ref{CollisionalMeasure} suggest that evaporative
cooling on pure samples of either $^{86}$Sr or $^{88}$Sr cannot be
very efficient in our experimental conditions. $^{86}$Sr presents
an extremely large elastic cross-section, a good point to
establish a fast thermalization, but the 3-body recombination rate
introduces a loss channel that is fatal with the typical
geometries accessible through optical dipole trapping. An optical
trap with a much larger trapping volume would partially suppress
this loss channel.\cite{Weber2002} $^{88}$Sr instead turns out to
be stable against 3-body decay, but the small elastic
cross-section results in a long thermalization time compared to
typical trap lifetime.

On the other hand, our results suggest a novel all-optical sympathetic
cooling scheme.\cite{Ferrari2006} In the isotope mixture the relatively
large inter-species cross-section results in thermalization times
typically of the order of few milliseconds. This thermalization is
fast even on the time scale of laser cooling on the
intercombination $^1$S$_0$-$^3$P$_1$ transition. Moreover, due to the
164\, MHz
$^{86}$Sr-$^{88}$Sr isotopic shift and the natural linewidth of 7.6\,kHz
for the $^1$S$_0$-$^3$P$_1$ transition, laser cooling on one
isotope has negligible effect on the other one. It
is then possible to cool sympathetically a dense and optically thick cloud
of one isotope (for instance
$^{88}$Sr) via optical cooling of a small sample of the other
isotope ($^{86}$Sr). Continuous laser cooling of $^{86}$Sr
provides heat dissipation in the sample, while the small optical
depth on $^{86}$Sr does not limit the achievable minimum
temperature. Sympathetic cooling with neutral atoms normally
requires a thermal bath with heat capacity large with respect to
that of the sample to be cooled. This is due to the fact that when
the thermal bath is cooled by evaporative cooling, each lost atom
carries an energy of the order of few times the temperature of the
sample. In the case of optical-sympathetic cooling, each
laser-cooled atom can subtract an energy of the order of the
optical recoil in a time corresponding to a few lifetimes of the
excited state, without being lost.

\begin{figure}[t]
\centerline{\psfig{file=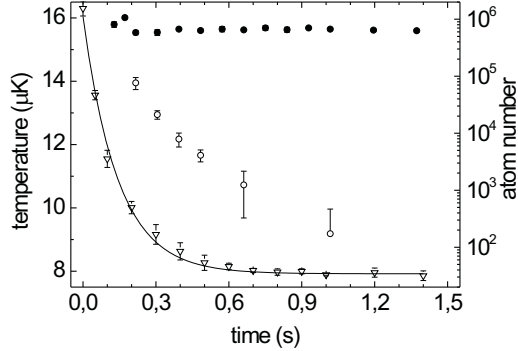,width=7cm}} 
\caption{Dynamics of an optically trapped $^{88}$Sr cloud
sympathetically cooled with laser cooled $^{86}$Sr. Filled
circles: $^{88}$Sr atom number. Open circles: $^{86}$Sr atom
number. The number of $^{88}$Sr atoms remains constant during the
process, while $^{86}$Sr decays exponentially with a 80\,ms time
constant. Under optimized conditions, the temperature (triangles)
decreases with a 150\,ms time constant and the mixture is always
at thermal equilibrium. Reprinted figure with permission from G.
Ferrari et al., Phys. Rev. A 73, 023408 (2006). Copyright (2006)
by the American Physical Society.}\label{SimpatheticCooling}
\end{figure}

We implemented the optical-sympathetic cooling scheme by extending
the temporal overlap between the FORT and the $^{86}$Sr red MOT
after switching off the $^{88}$Sr MOT. Figure
\ref{SimpatheticCooling} reports the dynamics of
optical-sympathetic cooling, starting from an initial temperature
of 15-20\,$\mu$K, limited by density dependent
heating.\cite{Poli2005,Weber2002} It can be observed that the
cooling process does not induce losses on $^{88}$Sr while the
number of $^{86}$Sr atoms exponentially decays with a 80\,ms
lifetime. About 100\,ms after switching off the $^{88}$Sr red MOT,
we observe that the mixture attains thermal equilibrium.

Under optimized conditions (overall optical intensity $100
I_{sat}$) the temperature decays from the initial value with a
150\,ms time constant. The minimum attainable temperature depends
both on the intensity of the $^{86}$Sr cooling beam, and the
$^{88}$Sr density. By keeping the cooling parameters on $^{86}$Sr
fixed at the optimum value and by varying the number of trapped
$^{88}$Sr, we determined the dependence of the final temperature
on the $^{88}$Sr density. For $6 \times 10^5$
$^{88}$Sr atoms trapped in the FORT, the final temperature is
6.7\,$\mu$K at a peak density of $1.3 \times 10^{14}$ cm$^3$; the
corresponding phase-space density is $5 \times 10^{-2}$. This
value is only a factor of two lower than what was
previously obtained without forced evaporation,\cite{Ido2000} but
it exhibits more favorable conditions for starting evaporative
cooling, considering the higher spatial density (more than one
order of magnitude higher) and the number of trapped atoms (gain
$2 \div 10$).

Indeed we applied an evaporative
cooling stage on $^{88}$Sr by reducing the FORT intensity after the optical sympathetic cooling.
This produced an
increase in phase-space density by roughly a factor of 4, giving a
maximum of $2 \times 10^{-1}$. Such result is consistent with a numerical simulation of
forced evaporation in our experimental conditions. The gain in phase-space density during evaporative
cooling is basically limited by the  $^{88}$Sr elastic cross-section. A promising way towards
 Bose-Einstein condensation seems to be the use of a dipole trap with variable
geometry, to compensate for the reduction in thermalization rate during
evaporation.\cite{Weiss2003}


\begin{figure}[t]
\centerline{\psfig{file=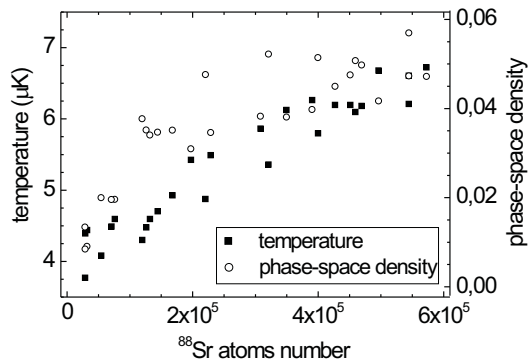,width=7cm}} 
\caption{Temperature and phase-space density of the $^{88}$Sr
cloud sympathetically cooled with laser cooled $^{86}$Sr, as a
function of the $^{88}$Sr atom number. Reprinted figure with
permission from G. Ferrari et al., Phys. Rev. A 73, 023408 (2006).
Copyright (2006) by the American Physical
Society.}\label{PhaseSpaceDensity}
\end{figure}

Figure \ref{PhaseSpaceDensity} shows the dependence of $^{88}$Sr
temperature after the optical sympathetic cooling (without
evaporative cooling)
 on the number of atoms
in the trap. We determine the density-dependent heating $dT/dn \simeq 2
\mu$K/($10^{14}$\,cm$^{-3}$), which is 20 times lower than the
equivalent value for pure laser cooled
$^{88}$Sr.\cite{KatoriMOT1999} This large reduction is a direct
consequence of the strong selectivity of the $^1$S$_0$-$^3$P$_1$
transition with respect to the two isotopes. The limit on the
$^{86}$Sr temperature of $4 \mu$K for zero $^{88}$Sr density can
be attributed to the laser cooling dynamics in the tightly
confining potential of the FORT.

\section{Ultracold Sr atoms as quantum sensors} \label{Microsensor}

Ultracold atomic strontium is particularly suited for applications
in the field of quantum sensors. Atom interferometry has already
been used with alkali-metals for precision inertial
sensing,\cite{Peters1999,Gustavson2000} for measuring fundamental
constants,\cite{Wicht2002,Clade2006,Stuhler2003} and testing
relativity.\cite{Fray2004} The extremely small size of ultracold
atomic samples enables precision measurements of forces at
micrometer scale. This is a challenge in physics for studies of
surfaces, Casimir effect,\cite{Antezza2005} and searches for
deviations from Newtonian gravity predicted by theories beyond the
standard model.\cite{Long2003,Dimopoulos2003,Samullin2005}

An interesting class of quantum devices is represented by
ultracold atoms confined in an optical lattice, that is a dipole
trap created by a laser standing wave.\cite{Bloch2005} In
particular, Bloch oscillations were predicted for electrons in a
periodic crystal potential in presence of a static electric
field\cite{Bloch1929} but could not be observed in natural
crystals because of the scattering of electrons by the lattice
defects. They were directly observed using atoms in an optical
lattice.\cite{Raizen1997}

The combination of the periodic optical potential and a linear
potential produced by a constant force $F$ along the lattice
wave-vector gives rise to Bloch oscillations at frequency $\nu_B$
given by $\nu_B = \frac{F \lambda_L}{2h}$, where $\lambda_L$ is
the wavelength of the light producing the lattice, and $h$ is
Planck's constant. Since $\lambda_L$ is well known, the force
along the lattice axis can be accurately determined by measuring
the Bloch frequency $\nu_B$. In order to perform a sensitive force
measurement, a long coherence time with respect to the measurement
duration is required. The most common effects limiting the
coherence time for ultracold atoms are perturbations due to
electromagnetic fields and atom-atom interactions. $^{88}$Sr is in
this respect an ideal choice because in the ground state it has
zero orbital, spin and nuclear angular momentum. This makes it
virtually insensitive to stray magnetic fields. In addition, as
shown in section \ref{CollisionalMeasure} $^{88}$Sr has remarkably
small atom-atom interactions. Such properties make Sr in optical
lattices a unique sensor for small-scale forces with better
performances and reduced complexity compared to proposed schemes
using degenerate Bose or Fermi gases.\cite{Anderson1998,Roati2004}
This enables new experiments on gravity in unexplored regions.

We tested such idea by measuring the gravitational acceleration in
our laboratory with a $^{88}$Sr sample in a vertical optical
lattice. To this end, we cool $\sim 5 \times 10^5$ atoms in the
red MOT down to the recoil temperature (see section \ref{RedMOT}),
so that the vertical momentum distribution is narrower than the
width of the first Brillouin zone.\cite{Bloch1929}. Then we
release the atoms from the MOT and we switch on adiabatically a
one-dimensional optical lattice.

The lattice potential is originated by a single-mode
frequency-doubled Nd:YVO$_4$ laser ($\lambda_L = 532$\,nm)
delivering up to 350\,mW on the atoms with a beam waist of $200
\mu$m. The beam is vertically aligned and retro-reflected by a
mirror producing a standing wave with a period
$\frac{\lambda_L}{2} = 266$\,nm. The corresponding photon recoil
energy is $E_R = \frac{h^2}{2m\lambda^2} = k_B \times 381$\,nK. We
populate about 100 lattice sites with $2 \times 10^5$ atoms at an
average spatial density of $\sim 10^{11}$\,cm$^{-3}$. We leave the
atoms in the optical lattice for a variable time $t$, then we
switch off the lattice adiabatically and we measure the momentum
distribution of the sample by time-of-flight imaging, after a free
fall of 12\,ms.

\begin{figure}[t]
\centerline{\psfig{file=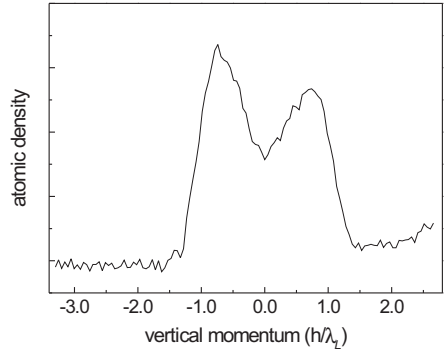,width=7cm}} 
\caption{Vertical momentum distribution of the atoms at the Bragg
reflection.}\label{TwoPeaks}
\end{figure}

We integrate along the
horizontal direction the optical thickness obtained by absorption imaging.
The resulting curve gives
the vertical momentum distribution of the atomic sample: in figure
\ref{TwoPeaks} we show a typical plot after the integration.

\begin{figure}[t]
\centerline{\psfig{file=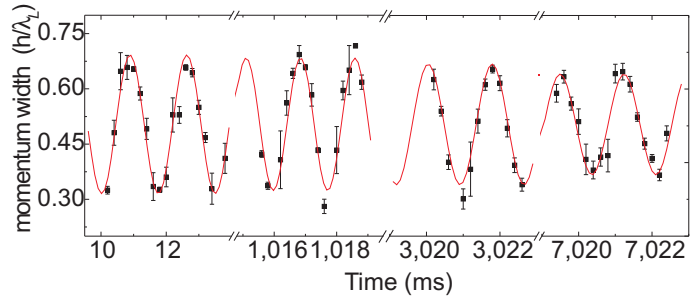,width=9cm}} 
\caption{Time evolution of the width of the atomic momentum
distribution, showing Bloch oscillation of $^{88}$Sr atoms in the
vertical 1-dimensional optical lattice under the effect of
gravity. From the the data fit, a Bloch frequency $\nu_B =
574.568(3)$\,Hz is obtained with a damping time $\tau\sim 12$\,s
for the oscillations.}\label{BlochOsc}
\end{figure}

We fit the measured momentum distribution  with the sum of two
Gaussian functions. From each fit we extract the vertical momentum
center of the lower peak and the width of the atomic momentum
distribution. We find that the latter is less sensitive against
noise-induced perturbations to the vertical momentum. We can
observe $\sim 4000$ Bloch oscillations in a time $t = 7$\,s (see
figure \ref{BlochOsc}), with a damping time $\tau\sim 12$\,s. To
our knowledge, the present results for number of Bloch
oscillations, duration, and the corresponding number of coherently
transferred photon momenta, are by far the highest ever achieved
experimentally in any physical system.

From the measured Bloch frequency $\nu = 574.568(3)$\,Hz we
determine the gravity acceleration along the optical lattice
$g_{meas} = 9.80012(5)$\,ms$^{-2}$.
The estimated sensitivity is $5 \times 10^{-6} g$. We expect that
such precision may be increased by one order of magnitude by using
a larger number of atoms, and reducing the initial temperature of
the sample.

\section{Frequency measurements on the Sr intercombination lines}

\label{1S3PFreqMeasure}

The advent of laser cooling techniques had important consequences
on time-frequency metrology, by reducing the uncertainty on the
frequency of atomic clocks caused by atom motion.

The recent progresses in two related fields, namely
high-resolution laser spectroscopy and direct optical-frequency
comb generation, opened the way to a new generation of frequency
standards based on transitions in the optical domain. The use of
frequency combs based on self-mode-locked femtosecond lasers has
made possible, for the first time, relatively simple
optical-frequency measurements.\cite{Udem1999,Diddams2000} On the
other hand, the realization of lasers with ultra-narrow emission
band now enables spectroscopy on forbidden optical transitions
with quality factor $Q = \frac{\nu}{\Delta\nu}$ in excess of
$10^{15}$.

Because of their higher frequency, optical transitions have the
potential for greatly improved accuracy and stability compared to
conventional atomic clocks based on microwave frequency
transitions.\cite{Udem2002} Different transitions are now
considered as optical-frequency standards, involving single ions
and neutral atoms.\cite{Udem2001,Gill2005} While single ions offer
an excellent control on systematic effects, clouds of laser cooled
atoms have the potential for extremely high precision, as the
large number of atoms reduces the quantum projection noise.
Perhaps the use of optical lattices at the magic wavelength to
confine neutral atoms in the Lamb-Dicke regime, as proposed for
the first time on strontium,\cite{KatoriFORT1999,Katori2003}
summarizes the best of both worlds; that is, a large number of
quantum absorbers with negligible shift of the optical clock
transition due to external fields, Doppler effect and collisions.

Among the neutral atoms, Sr has long been considered as one of the
most interesting candidates.\cite{Hall1989} Several features, some
of which are specific to this atom, allow different possibilities
for the realization of a high precision optical clock. The visible
intercombination 5$^1$S-5$^3$P lines from the ground state are
easily accessible with semiconductor lasers. Depending on the
specific fine-structure component and on the isotope,  a wide
choice of transitions with different natural linewidths is
possible (see section \ref{Strontiumatom}).

The first phase-coherent absolute frequency measurement of a Sr
intercombination line was performed by our group on the
5$^1$S$_0$-5$^3$P$_{1}$ transition, using a thermal atomic
beam.\cite{Ferrari2003} This represented an improvement by several
orders of magnitude with respect to previous data.\cite{Tino1992}
Our result is in agreement with subsequent phase-coherent
measurements performed by the BNM-SYRTE group on a thermal
sample,\cite{Courtillotclock} and by the JILA group on a
free-falling ultracold sample using a red MOT.\cite{Ido2005} The
JILA measurement provided a further improvement in the accuracy by
more than two orders of magnitude.

However, the $^1$S$_0$-$^3$P$_{1}$ transition is not best suited
as a final frequency reference, due to its natural linewidth of
7.6\,kHz. Some groups recently began working on the 698\,nm
$^1$S$_0$-$^3$P$_{0}$ line, that is strictly forbidden in the even
isotopes since it is a $J=0\to J=0$ transition. In $^{87}$Sr the
hyperfine mixing enables direct $^1$S$_0$-$^3$P$_{0}$ excitation
with a transition probability of $\sim 1 $\,mHz. The BNM-SYRTE
group first measured such transition in a blue MOT with an
uncertainty of 15\,kHz.\cite{Courtillotclock} The Tokyo group
performed the first absolute frequency measurement in an optical
lattice on this transition,\cite{Takamoto2005} followed by the
JILA and BNM-SYRTE groups.\cite{Ludlow2005,Targat2006} All groups
estimated an uncertainty $\leq 20$\,Hz for the absolute transition
frequency. The corresponding 578\,nm $^1$S$_0$-$^3$P$_{0}$
transition in Ytterbium was observed at NIST in a $\sim 70 \mu$K
sample on the two odd isotopes, $^{171}$Yb and $^{173}$Yb, with an
uncertainty of $\sim 4$\,kHz.\cite{Hoyt2005}

In spite of its large quality factor, the $^1$S$_0$-$^3$P$_{0}$
transition in odd Sr and Yb isotopes suffers from residual
sensitivity to stray magnetic fields and optical lattice
polarization, besides a complex line structure due to the presence
of many magnetic sublevels. Several groups are now looking at the
even Sr or Yb isotopes as the best candidates to represent optical
frequency standards based on neutral atoms. Some groups have
proposed different methods to directly excite the clock transition
on the even Sr or Yb isotopes, by properly engineering the atomic
level structure to create a finite and controllable
$^1$S$_0$-$^3$P$_{0}$ transition probability. These methods
basically consist in coupling the metastable $^3$P$_{0}$ level to
other electronic states by using either multiple near-resonant
laser fields,\cite{Santra2005,Hong2005} or simply a small static
magnetic field.\cite{Taichenachev2005} The latter scheme has been
experimentally demonstrated on $^{174}$Yb at
NIST.\cite{Barber2005} The possible instability due to
site-to-site tunneling in optical lattice clocks has been
addressed by the BNM-SYRTE group.\cite{Lemonde2005} For accuracy
goals at the $10^{-18}$ level they propose the use of vertical
optical lattices, in order to lift the degeneracy between adjacent
potential wells.

\subsection{Frequency measurement on the $^1$S$_0$-$^3$P$_1$
transition with a thermal beam} \label{OpticalFreqMeasure}

In this section we discuss our precision frequency
measurements on the intercombination 5$^1$S$_0$-5$^3$P$_{1}$
transition.\cite{Ferrari2003} Using a femtosecond laser comb, we
determined the absolute frequency of the transition for $^{88}$Sr
and $^{86}$Sr and an accurate value for the isotope shift.

The frequency measurements have been performed through saturation
spectroscopy on a thermal beam with the apparatus described in
appendix \ref{RedLaser}. As frequency-comb generator we employed a
commercial system based on a Kerr-lens mode-locked Ti:Sa laser
with a repetition rate of 1\,GHz.\cite{Cundiff2001} Its repetition
rate and carrier envelope offset frequency were locked to a GPS
stabilized quartz oscillator. The strontium atomic beam is
obtained from the metal heated to 830\,K. Residual atomic beam
divergency is 25\,mrad and the typical atomic density in the
detection region is 10$^8$ cm$^{-3}$. We derive a Doppler-free
signal using a retro-reflected laser beam perpendicular to the
atomic beam. The fluorescence light from the laser excited atoms
is collected on a photomultiplier tube.

 We estimate the indetermination on the reflection angle
of the laser beam to be less than 10\,$\mu$rad. The peak beam
intensity of 60\,$\mu$W/cm$^2$ (to be compared to the saturation
intensity of 3\,$\mu$W/cm$^2$) was chosen to obtain sufficient
signal-to-noise for the RC lock onto the atomic resonance. A
uniform magnetic field of 10\,G (see figure \ref{689Setup})
defines the quantization axis in the interrogation region such
that the light is $\pi$ polarized. The double pass AOM next to the
atomic detection (AOM3) is frequency modulated at 10\,kHz to
derive the locking signal of the cavity onto the atomic line.

\begin{figure}[t]
\centerline{\psfig{file=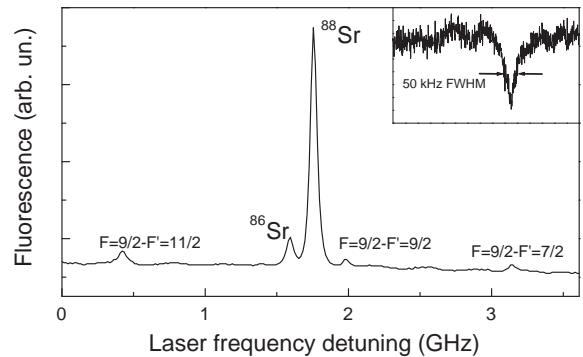,width=9cm}} 
\caption{Fluorescence spectrum of the strontium
$^1$S$_0$-$^3$P$_1$ line at 689\,nm. The lines of the two bosonic
isotopes $^{86}$Sr and $^{88}$Sr, together with the hyperfine
structure of the fermionic $^{87}$Sr, can be resolved. The
linewidth corresponds to the residual first order Doppler
broadening in the thermal beam. Inset: sub-Doppler resonance of
$^{88}$Sr recorded by saturation spectroscopy using two
counterpropagating laser beams. The amplitude of the dip is 10\,\%
of the Doppler signal. Reprinted figure with permission from G.
Ferrari et al., Phys. Rev. Lett. 91, 243002 (2003). Copyright
(2006) by the American Physical Society.}\label{RedFluor}
\end{figure}

Figure \ref{RedFluor} shows the Doppler broadened resonances of
$^{88}$Sr, $^{86}$Sr, and the hyperfine structure of $^{87}$Sr.
The residual atomic beam divergency produces a residual Doppler
broadening of 60\,MHz FWHM. In the inset, the sub-Doppler signal
for $^{88}$Sr is shown. Two independent measurements of the
sub-Doppler resonance show a FWHM of about 50\,kHz, which is in
agreement with the expected value considering the saturation and
transit time broadening, and the recoil splitting.

\begin{figure}[t]
\centerline{\psfig{file=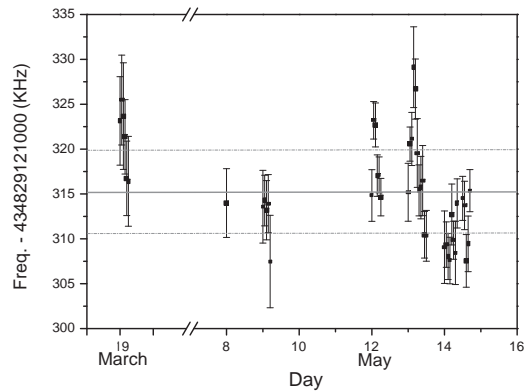,width=8cm}} 
\caption{Chronological plot of the optical frequency measurements.
The error bars correspond to the standard deviation for each data
set. Reprinted figure with permission from G. Ferrari et al.,
Phys. Rev. Lett. 91, 243002 (2003). Copyright (2006) by the
American Physical Society.}\label{FreqencyMeasure}
\end{figure}

Figure \ref{FreqencyMeasure} shows the result of the measurement
of the $^{88}$Sr transition frequency taken over a period of
several days. Each data point corresponds to the averaging of the
values resulting from consecutive measurements taken with a 1\,s
integration time over $100 \div 200$\,s. The error bars correspond
to the standard deviation for each data set. We evaluated first-
and second-order Doppler and Zeeman effects, ac Stark shift,
collisional shifts, and mechanical effects of light. The resulting
value for the $^{88}$Sr transition frequency, including the
corrections discussed previously, is 434 829 121 311 (10)\,kHz,
corresponding to a 1$\sigma$ relative uncertainty of $2.3 \times
10^{-11}$.

\subsection{$^{86}$Sr - $^{88}$Sr isotopic shift measurement} \label{IsotopeShift}

With a minor change in the apparatus, we locked simultaneously the
frequency of two laser beams to the sub-Doppler signals of
$^{86}$Sr and $^{88}$Sr. This system allowed us to measure the
isotopic shift of the $^1$S$_0$-$^3$P$_1$ transition by counting
the beat note between the two interrogating beams. For this
purpose, the reference cavity is locked to the $^{88}$Sr resonance
as described previously and the light for $^{86}$Sr is derived
from the same laser beam and brought to resonance through AOMs.
The two beams are overlapped in a single mode optical fiber and
sent to the interrogation region. By frequency modulating the
beams at different rates and using phase sensitive detection we
get the lock signal for both the isotopes from the same
photomultiplier. In this isotope-shift measurement most of the
noise sources are basically common mode and rejected. The measured
$^{88}$Sr-$^{86}$Sr isotope shift for the $^1$S$_0$-$^3$P$_1$
transition is 163 817.4 (0.2)\,kHz. This value represented an
improvement in accuracy of more than three orders of magnitude
with respect to previously available data.\cite{Buchinger1985} The
$^{86}$Sr optical frequency then amounts to 434 828 957 494
(10)\,kHz.

\section{Conclusions}
\label{Conclusion}

The strontium atom is an attractive candidate both for physical
studies and for applications. We have shown how the properties of
atomic strontium  are suited for laser cooling and trapping, for
the study of ultracold atomic physics, and for applications to
optical frequency metrology or to micrometric force sensors.

Future work on ultracold strontium offers intriguing perspectives,
including the possible realization of a nearly non-interacting
Bose-Einstein condensate, the realization of an optical clock with
ultimate stability, or direct experimental tests of theories
beyond the standard model.

\appendix

\section{Experimental setup}
\label{ExperimentalSetup}

The typical experimental setup for Sr laser cooling and trapping
basically includes a vacuum system, a blue laser source for the
atom collection and first cooling stage, a red laser for second
stage cooling and precision spectroscopy, and an infrared laser
source for the optical dipole trap. In the following we illustrate
these main parts as they are realized in our laboratory.

\subsection{Vacuum system} \label{VacuumSystem}

The apparatus consists in three major parts: the oven, the Zeeman
slower, and the MOT chamber. The oven generates an atomic
strontium vapor by sublimation from metallic strontium kept at
$\sim 800$ Kelvin. The vapor is collimated into an atomic beam
passing through a nozzle filled with about 200 stainless steel
capillaries 8\,mm long which insure a ballistic divergence of the
atomic beam of 20\,mRad.\cite{Courtillot2003} Keeping the
capillaries at a higher temperature prevents internal vapor
condensation and clogging.

\begin{figure}[t]
\centerline{\psfig{file=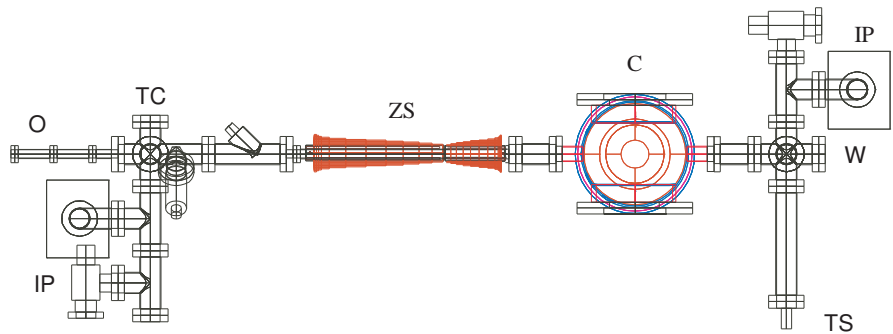,width=12cm}} \vspace*{8pt}
\caption{Vacuum aparatus. O: oven for St sublimation; IP: ion
pumps; TC: cell for transverse cooling; ZS: differential pumping
tube for the Zeeman slower; C: MOT cell; TS: titanium sublimation
pump; W: sapphire window for the Zeeman slower laser
beam.}\label{VacuumPicture}
\end{figure}

Following the atomic trajectory (see the sketch in figure
\ref{VacuumPicture}), the atoms pass through a region with radial
optical access for 2D transverse cooling,\cite{Shimizu,Rasel1999}
they are decelerated through the Zeeman slower,\cite{Prodan1982}
and finally stopped in the cell that hosts the MOT. The optical
access for for 2D transverse cooling is provided by a CF35 cube
aligned on the atomic beam. On the two free direction two pairs of
windows, AR coated for 461 and 689\,nm, are sealed with modified
copper gaskets.\cite{WindowSeal} A differential pumping stage
between the transverse cooling region and the Zeeman slower
insures decoupled background pressure between the oven and MOT
region. The oven is pumped by a 20\,l/s ion pump while the MOT is
pumped by a 20\,l/s ion pump and a titanium sublimation pump. With
this setup, in operation condition, we achieve a pressure of
$10^{-7}$ Torr in the transverse cooling region, and a pressure of
$10^{-9}$ Torr on the MOT cell. A BK7 window on the atomic beam
axis provides access for the Zeeman slower decelerating beam. To
prevent chemical reaction of strontium and darkening, the
anti-reflection is only present on the outer side of the window ,
and the window is heated to $\sim 450$\,Kelvin. Paying attention
to block the Sr beam when unnecessary, the window presents a dim
shadow after two years of operation. In the future we plan to
exchange the window with a sapphire one, again anti-reflection
coated on the outer side, from which deposited Sr can be easily
removed.

\begin{figure}[t]
\centerline{\psfig{file=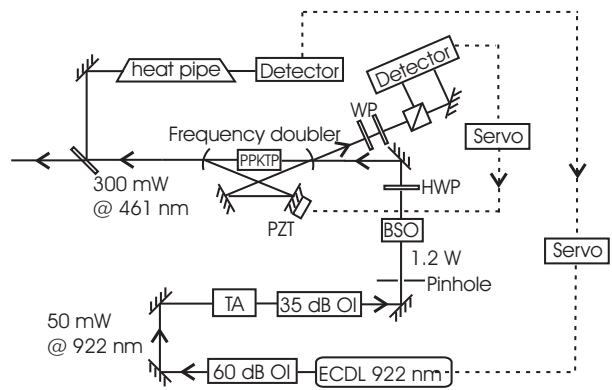,width=8cm}} \vspace*{8pt} \caption{Scheme of the laser at 461
nm. A distributed feedback laser (DFB) is amplified in a semiconductor tapered amplifier (TA), and
frequency doubled on a periodically poled KTP crystal. The non-linear conversion is improved by
placing the KTP crystal into an optical resonator. Optical isolators (OI) between the DFB and TA,
and between the TA and the frequency doubling stage. Solid lines represent the optical path,
dashed lines represent electrical connections. BSO: beam shaping optics.}\label{BlueSetup}
\end{figure}

\subsection{Blue laser sources} \label{BlueLaser}

The light at 461\,nm (see figure \ref{LevelScheme}) used for the
atomic beam deceleration and capture in the MOT is produced by
second-harmonic generation (SHG) of a 922\,nm semiconductor laser
(see figure \ref{BlueSetup}). We generate the 922 nm radiation
with a master oscillator-parametric amplifier system (MOPA). An
anti-reflection coated laser diode mounted in an extended cavity
in Littrow configuration (ECDL)\cite{Littrow1991} delivers 50\,mW
at 922 nm, that is amplified to 1.2 W in a tapered amplifier (TA).
Optical isolators are placed between the ECDL and the TA, and
between the TA and the frequency doubler cavity, in order to
prevent optical feedback into the master laser, and optical
damages on the amplifier. The frequency doubler is composed of a
20\,mm long periodically-poled KTP crystal, placed in an optical
build up cavity. The crystal facets are anti-reflection coated
both at 922 and 461\,nm (R\,$<$\,0.2\,\%) and the poling period is
chosen to fulfill quasi-phase matching of our wavelength at room
temperature. The resonator has an input coupling mirror with
11\,\% transmission and it is held in resonance with the input
light feeding the error signal from a H\"{a}nsch-Couillaud
detector\cite{Couillaud81} to a PZT controlled folding mirror.
Under optimal conditions we obtain 300\,mW in the blue and
routinely we work with 200\,mW. We frequency stabilize this blue
source to the $^1$S$_0$-$^1$P$_1$ line of $^{88}$Sr by means of
conventional saturated spectroscopy in a strontium heatpipe. The
servo loop acts on the piezo of the ECDL.

The light used for atomic manipulation and detection is brought to
the vacuum system through single-mode polarization-maintaining
optical fibers to improve the long-term beam pointing stability.

As we discussed in section \ref{OpticalPumping}, we employ a
turquoise laser at 497\,nm to increase the MOT lifetime and number
of atoms loaded. As in the case of 461\,nm source, there are no
laser diodes available at 497\,nm and the simplest method to
produce this light is frequency doubling an IR laser at 994\,nm.
In this application 1\,mW of light is sufficient to saturate the
process, then we do not need any amplification of the IR light
before frequency doubling. The source at 497\,nm differs from that
at 461\,nm in few parts. The master laser is an anti-reflection
coated diode in ECDL Littrow configuration. After the beam shaping
optics, the IR light is coupled into a bow-tie cavity that
contains a 17\,mm long, b-cut potassium niobate crystal which is
kept at 328\,K in order to satisfy non-critical phase matching for
SHG. The crystal facets are AR coated both for the fundamental and
the blue light and, like the 461\,nm source, the cavity is kept
resonant to the 994\,nm laser with a H\"{a}nsch-Couillaud locking.
Since this laser operates among Sr excited states it is not
possible to lock the laser to the atomic line on a simple
heatpipe. Possible frequency stabilization methods include locking
to a reference cavity or spectroscopy on an atomic sample with a
suitable fraction of excited atoms, such as a hollow cathode lamp,
or an heatpipe with gas discharge, or simply the MOT. We chose the
latter system, in spite of the fact that the blue MOT fluorescence
signal is not continuously available during our measurement
cycles. In fact the short-term laser frequency stability is
sufficient to keep it in resonance with the $^3$P$_2$-$^3$D$_2$
transition for some tens of seconds. Thus we leave the laser free
running, and we manually adjust its frequency at the beginning of
the measurement cycle by acting on the piezo of the ECDL to
maximize the MOT fluorescence.

\subsection{Red laser source} \label{RedLaser}

\begin{figure}[t]
\centerline{\psfig{file=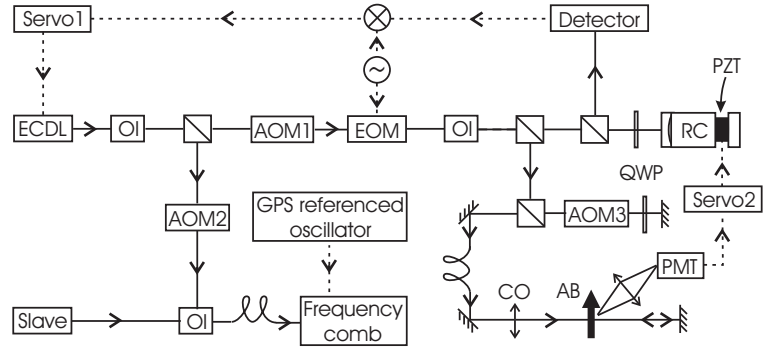,width=10cm}} \vspace*{8pt}
\caption{Scheme of the 689\,nm laser. Experimental setup used for
the frequency measurement on the Sr intercombination line. Optical
isolators (OI) and acusto-optic modulators (AOM) eliminate
feedback among the master laser (ECDL), the slave laser, the
electro-optic modulator (EOM) and the reference cavity (RC).  QWP:
quarter wave-plate. CO: collimation optics. PMF: polarization
maintaining fiber. PMT: photo multiplier tube. The same apparatus
is used for Sr second stage cooling and trapping, but the thermal
beam is replaced by an heat-pipe.}\label{689Setup}
\end{figure}

The 689\,nm source is composed of a laser diode frequency-locked
to an optical cavity whose modes are stabilized to keep the laser
on resonance with the atomic line. A scheme of the experimental
setup is given in figure \ref{689Setup}. An extended cavity
laser-diode mounted in the Littrow configuration delivers 15\,mW
at 689 nm. Optical feedback into the ECDL is prevented by a 40\,dB
optical isolator and a single pass acusto-optic modulator in
cascade. The laser linewidth is reduced by locking the laser to an
optical reference cavity (RC) with standard Pound-Drever-Hall
technique;\cite{Pound-Drever-Hall} the phase modulation is
produced by an electro-optic modulator (EOM) driven at 11\,MHz and
leaves 85\,\% of the power in the carrier. The reference cavity
has a free spectral range (FSR) of 1.5\,GHz and a finesse of $\sim
7000$. On one side of the quartz spacer we glued a concave mirror
(R$=50$\,cm) while on the other side a PZT is glued between the
spacer and a flat mirror in order to steer the modes of the cavity
by more than one FSR.

The locking loop includes a low frequency channel acting on the
PZT of the ECDL (1\,kHz bandwidth), and a high frequency channel
acting on the laser-diode current supply ($\sim 3$\,MHz
bandwidth). Under lock condition more than 55\,\% of the incident
light is transmitted through the cavity. From the frequency noise
spectrum obtained by comparison with an independent cavity we can
infer a laser linewidth lower than 20\,Hz, and more than 90\,\% of
the optical power in the carrier.\cite{Ferrari2004,Poli2006} The
RC is acoustically isolated, though we do not keep it under
vacuum.\cite{UltraStableCavity} The optical table is actively
isolated from seismic noise with pneumatic legs. The frequency
drifts of the cavity are compensated by the servo to the atomic
signal which acts on the PZT of the high finesse cavity with a
20\,Hz bandwidth. In the frequency measurement experiment
described in section \ref{FreqencyMeasure} the Doppler free
saturated fluorescence on a thermal strontium beam provides the
signal for cavity stabilization on the atomic line. The thermal
beam source has a similar design as described in
\ref{VacuumSystem} and it is pumped by a 20 l/s ion pump. However,
when using the stable 689\,nm laser for second stage cooling and
trapping, as described in section \ref{RedMOT}, we employ an
heat-pipe kept at $\sim 750$\,K instead of the thermal beam, thus
obtaining a larger signal and a more robust lock.

\subsection{Infrared laser sources for dipole trapping} \label{IRlaser}

\begin{figure}[t]
\centerline{\psfig{file=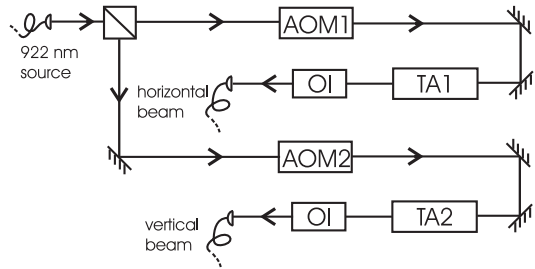,width=7cm}} 
\caption{The 922\,nm laser source for the optical dipole trap.
}\label{FORTsources}
\end{figure}

After the production of an ultracold sample in double stage
magneto-optical trapping, we transfer the atoms in an optical
dipole trap made of two infrared laser beams crossing each other
near the waist. The two beams, respectively aligned along the
horizontal and vertical direction, are produced with two
independent TA injected with light coming from the same infrared
source used for producing the blue light (see figure
\ref{FORTsources}), that is close enough to the ``magic wavelength'' for the $^1$S$_0$-$^3$P$_{1}$
transition.
Typically 300\,mW of
light coming from that source are coupled into a fiber and sent to the two amplifiers.
The injection of the two TAs it is regulated by two AOMs which are
used to shift the frequency of the two beams (avoiding
interference at the center of the dipole trap) and to apply a fast
control to the output TA optical power. For mode cleaning and
delivering the output beams from the TAs to the atoms, we use two
independent single mode fibers. Typically we obtain about 650\,mW
and 440\,mW at the fiber output for the horizontal and vertical
beams, respectively. We finally focus the beams at the MOT center,
with 1/e$^2$ radii of 15 $\mu$m and 35 $\mu$m respectively.

\end{document}